# MIMO CDMA-based Optical *SATCOMs*: A New Solution

Makan Zamanipour
IEEE Member
Tehran, Iran
makan.zamanipour.2015@ieee.org

Mohammadali Mohammadi
Shahrekord University
Shahrekord, Iran
m.a.mohammadi@eng.sku.ac.ir

*Abstract*— A new scheme for MIMO CDMA-based optical satellite communications (O*SATCOMs*) is presented. Three independent problems are described for up-link and down-link in terms of two distinguished optimization problems. At first, in up-link, Pulse-width optimization is proposed to reduce dispersions over fibers as the terrestrial part. This is performed for return-to-zero (RZ) modulation that is supposed to be used as an example in here. This is carried out by solving the first optimization problem, while minimizing the probability of overlapping for the Gaussian pulses that are used to produce RZ. Some constraints are assumed such as a threshold for the peak-to-average power ratio (PAPR). In down-link, the second and the third problems are discussed as follows, jointly as a closed-form solution. Solving the second optimization problem, an objective function is obtained, namely the MIMO CDMA-based satellite weight-matrix as a conventional adaptive beam-former. The Satellite link is stablished over flat un-correlated Nakagami-*m*/Suzuki fading channels as the second problem. On the other hand, the mentioned optimization problem is robustly solved as the third important problem, while considering inter-cell interferences in the multi-cell scenario. Robust solution is performed due to the partial knowledge of each cell from the others in which the link capacity is maximized. Analytical results are conducted to investigate the merit of system.

## I. INTRODUCTION

Integrated satellites and terrestrial networks have been widely examined in the literature, [1]. In this study, two communication links are separately considered, namely up-link over the optical fiber cable as the terrestrial part, and down-link over the satellite link. We focus on the following aspects in this paper:

1) In up-link: Pulse-width optimization is proposed to reduce dispersions in optical fibers.

2) In down-link: The satellite weight-matrix is solved as a closed-form solution. As the second problem, this is accomplished in flat un-correlated Nakagami-m/Suzuki fading channels whose distributions are more complicated compared to Rayleigh fading channels.

3) Again in down-link, the mentioned weight-matrix is robustly obtained, considering inter-cell interferences, while no cooperation is attended for cells. I.e., cells have no knowledge of the others as the imperfect channel state information (ICSI) at the transmitter.

To clarify, at first, challenges and motivations are given as follows. Then, the contributions are given related to the discussed motivations to solve the mentioned challenges. Finally, related works are discussed.

At first, the motivations related to pulse-width optimization are discussed as follows. In optical satellite communications (OSATCOMs), one of the conventional ways to transfer datas is optical fiber cables in terrestrial parts. One of the main challenges in optical fibers is *dispersion issue*. Different modes of dispersions are well-known caused by many reasons which are not supposed to be discussed in here. Indeed, this is discussed as a different view. The total dispersion in fibers is realized as the *root mean square* of these dispersion types. Total dispersions make the *inter symbol interference* (ISI) as *pulse spreading*. In here, ISI is minimized, while considering the probability of overlapping for the Gaussian pulses that are used to produce RZ. For instance, this probability for non-RZ is higher than RZ. See next sections. It should be noticed that RZ as an example is considered and this method can be easily generalized to the other modulations such as on-off-keying. Pulse-width optimization to solve ISI is explored as the first optimization problem.

The challenges/motivations related to robust weighing over special fading channels are discussed as follows.

On the one hand, this is formulated for the used MIMO CDMA-based satellite link over flat un-correlated Nakagami-*m*/Suzuki fading channels. It is difficult to obtain e.g. the probability of error over the fading channels whose distributions are not Rayleigh. Indeed, the elements of the channel matrices are not zero-mean and unit-variance. Thus, the related equations are more complicated.

On the other hand, the design problem is more complicated, where the system is operated in the multi-cell scenario in which no cooperation is considered for cells. In other words, in here, no cell has the knowledge of the others. Hence, the design problem should be robustly investigated. Indeed, when the estimation of a parameter such as channels has any error as a more real and hard scenario, robust solution should be used. Moreover, this is performed in the multi-cell scenario as a real scenario.

Both the above-mentioned works are carried out as the second optimization problem in which the link capacity is maximized, while constraining some factors such as the transmit-power.

The contributions related to the first problem are given as follows. An approach is investigated to optimize the modulated pulse-width, so as to manage ISI, while solving the optimization problem (a) in which some constraints such as PAPR should be considered. This work has not been performed until now. Unlike previous studies that are given in next paragraphs, apart from the used circuits, a novel mathematical calculation is generally proposed in this study. In other words, the used pulse-width is precisely computed, while being no necessity to consider general aspects such as the cable length. This is actualized as a trade-off, when the increase in PAPR is acceptable.

The contributions related to the second problem are given. A closed-form solution is investigated for receive-powers over Nakagami-$m$/Suzuki fading channels that has not been investigated for SATCOMs until now. This is fulfilled, when exploiting the *Lagrange dual-decomposition method* for the convex optimization problem (b). Thereby, the problem is prepared to be used by the Lagrange dual-decomposition or the *eigen-value decomposition* methods. Indeed, the design problem is analyzed in more real and hard conditions. Furthermore, this is extended more in details. This is more helpful in correlated fading channels. See next sections for more discussion.

The contributions related to the third problem are given as follows. The used robust optimization is discussed as a new view, namely considering the worst case of interference powers as a constraint. In other words, this is examined as a proposition, *without loss of optimality*. Moreover, this is compared to the other works.

About previous works, related to the first problem, in [2], a fixed pulse-width was pointed out, considering the rate and the modulation agility. A tunable pulse-width management was proposed in [3], according to tuning the captured voltage on a phase modulator, jointly using a tunable dispersion element. A tunable RZ pulse generator based upon spectral line-by-line pulse shaping was introduced in [4].

Related to the second problem, despite [5, 6], the capacity for multi-beam satellites over Rician fading channels was investigated in [7].

Related to the third problem, the used approach in the robust optimization problem has not been performed until now. In [8], another solution was examined, whereas in here, this is performed according to the other *matrix norm features*. Meanwhile, the amount of the convergence for optimum values between this work and [8] is finally compared.

To summarize, the other differences of this study compared to previous works are discussed as follows. The joint proposed methods and the considered integrated scheme are completely different from the other works related to OSATCOMs CDMA-based satellite links under ICSIs. The proposed non-cooperative cellular SATCOM has not been discussed until now. In [9], the interferences on multi-cell satellites was non-robustly explored as an experimental study to allocate frequencies. In [10], a study was performed in different cases of the estimation error in multi-beam satellites, while proposing no closed-form solution. That was examined under the ICSI at the receiver. In [11], an adaptive space-time signal processing to cancel interferences, so as to detect users in CDMA-based SATCOMs was proposed, while reusing frequencies. See e.g. [12] for more comparison.

The remainder of the paper is organized as follows. The used MIMO CDMA-based optical SATCOM is stated in section II. In section III.A, up-link over fiber cables is discussed in which ISI and pulse-width optimization are examined, solving the optimization problem (a). Robust design to weight the non-cooperative cellular satellite link over non-Rayleigh fading channels is formulated in down-link in section III.B, solving the optimization problem (b). Some criteria such as the probability of error is used to evaluate the performance in section IV as the analytical results. Conclusions and proofs are given in section V and Appendix, respectively.

Notations: Matrices, vectors and scalars are denoted by boldfaced upper-case characters, boldfaced lower-case characters and non-boldfaced lower-case characters, respectively. $\mathbf{I}_M$ represents the M-by-M identity matrix. The operations $\|.\|_F$, $E(.)$, $Tr(.)$, $(.)^H$, $vec(.)$ and $\otimes$ are the Frobenius norm, the statistical expectation, the trace, the Hermitian, the vectorization operations and the Kronecker product. The dimension of MIMO channels is $M$.

## II. LINK DESCRIPTION

The optical satellite link that is considered in the multi-cell scenario, is illustrated as Fig. 1.

The *1*th channel refers to the main MIMO channel between the MIMO transmitter and the MIMO receiver in the related cell. Subsequently, the *2*th channel is related to inter-cell interference signals between neighbor cells. Subscripts 1 and 2 in all the equations such as the matrices $\mathbf{H}_{1,a}$ and $\mathbf{H}_{2,a}$ of sizes $M \times M$ (in the *a*th cell), refer to this point throughout the paper.

## III. PROBLEM FORMULATION

*A. Up-link: Optical Fiber Cable*

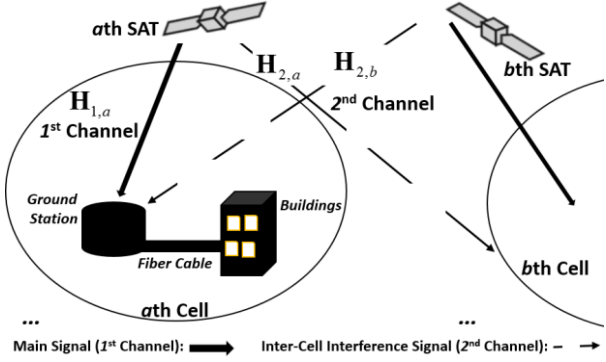

Fig. 1. The used optical satellite link in the multi-cell scenario

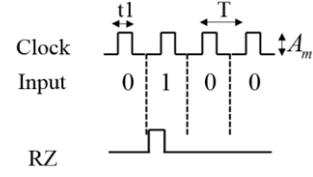

Fig.2. RZ modulation based on the examples for the clock and the input

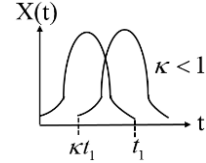

Fig.3. Overlapping for neighbor Gaussian pulses (related to the clock)

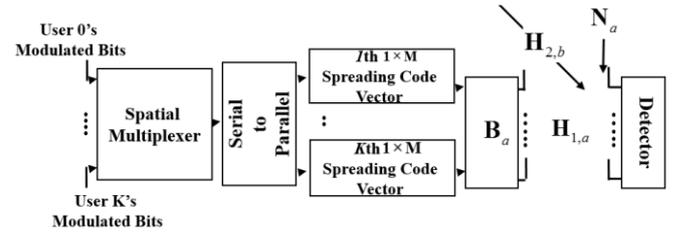

Fig.4. The satellite link in down-link

How to modulate as RZ, based upon the examples for the clock and the input signal is declared in Fig. 2. As shown, the main challenge is to reduce the pulse-width of the considered clock for RZ.

Suppose that each pulse of the clock is produced as a Gaussian pulse for the fiber cable. Let (1) be the optimization problem (a):

$$\text{a) } \min_{\kappa, t1} \quad \text{Pr}_{overlap} = \int_{\kappa t1}^{t1} pdf(x(t))dt$$

Subject to:

$$\text{b1) } PAPR: 10\log_{10}\frac{T}{t1}(dB) \leq PAPR_{th} \quad (1)$$

$$\text{b2) } OSNR: \frac{1}{\partial^2_{n,fiber}}\|\mathbf{H}_{fiber}\|_F^2\left\{\frac{1}{T}\int_0^T x^2(t)dt = \frac{t_1 A_m^2}{T}\right\} \geq OSNR_{tar}.$$

The probability of overlapping (1.a) for these pulses as the *cost function* should be minimized. $\kappa$ (as an arbitrary constant value that is smaller than 1) and $t1$ in Fig. 3 are supposed to be achieved as the *objective functions*. $PAPR_{th}$ as a threshold for PAPR for the clock that shown as an example in Fig. 2, should be defined as (1.b.1). $OSNR_{tar}$ as the *target* optical signal to noise ratio (*OSNR*) at the end of the fiber cable in up-link is written as (1.b.2) as the second constraint. $\mathbf{H}_{fiber}$ and $\partial^2_{n,fiber}$ refer to the fiber channel and the noise variance in the fiber cable. $pdf(x(t))$ is the Gaussian probability density function of the Gaussian pulse $x(t)$. $pdf(x(t))$ is a function with a mean and a variance that are unable to be used in the optimization problem as non-controllable parameters. How to have the mentioned overlapping for two neighbor Gaussian pulses is demonstrated in Fig. 3.

*B. Down-link: Satellite Link*

The transceiver in down-link is shown in Fig. 4, where $\mathbf{B}_a$ is the weight-matrix of size $M \times M$ related to the $a$th cell. Certainly, spreading codes and de-spreading codes have no effect on the equivalent channel norm. The spreading code vector is applied to the transmitted signal after being spatial multiplexed and *serial-to-parallel* in the used CDMA system. Likewise, users in here can be virtual such as applications. Subsequently, in (2) that is written for the $a$th cell, $\mathbf{H}_{2,b}$ are the channels between the other satellites (related to the other cells) and the $a$th cell:

$$\mathbf{Y}_a = \mathbf{X}_a \odot \left\{ \cdots \odot \left[\sum_{\substack{b=1\\b\neq a}}\cdots\right]\right\} \quad (2)$$

$\mathbf{N}_a$ and $\mathbf{X}_a$ are the sky noise matrix and the *atmospheric impact* matrix, respectively. $\odot$ and $Y_a$ are the Hadamard product and the received signal, respectively. Let $\mathbf{X}_a$ be identity for simplicity that refers to *high cross-polarization discrimination results* related to *cross-polarization impacts* as [13]. See [13] for more discussion. The first term in (2), defines the received main signal. The second term specifies the total interference.

$SNR_a$ is obtained as $Tr\{\tilde{\mathbf{b}}_a \mathbf{D}_a\}$ that should be multiplied by rain effects. See Appendix A. Applying rain effects is discussed in next paragraphs. $\mathbf{D}_a$ is related to means and variances of Nakagami-$m$ fading channels in the $a$th cell.

*Proposition 1*: *The worst case of the inter-cell interference power as* $\max_{\|\Delta\|_F \leq \xi}\{P_{Inter-cell}(\Delta)\}$ *for the $a$th cell caused by the $a$th SAT under the other cells, can be robustly written as (3). (3) can be written as (proof: see Appendix B):*

$$P_{\text{inter-cell}}(\xi) = Tr\{\tilde{\mathbf{b}}_a\left\{\tilde{\mathbf{H}}_{2,a}\tilde{\mathbf{H}}_{2,a}^H + \sqrt{M}\xi(\sqrt{M}\xi + 2\|\tilde{\mathbf{H}}_{2,a}\|_F)\mathbf{I}_M\right\}\}. \quad (3)$$

where $\xi$ is a parameter related to the robust optimization, i.e., the error matrix $\Delta$ related to the estimation of $\mathbf{H}_{2,a}$ should be constrained.

The optimization problem (b) for the $a$th cell is realized as (4) to achieve $\tilde{\mathbf{b}}_a$ as $vec(\mathbf{B}_a^H \mathbf{B}_a)$ at the transmitter:

$$\begin{aligned} a) & \max_{\tilde{\mathbf{b}}_a, a=1,...,b,...,A} \log_2(1+10^{\frac{-A_{R_a}}{10}} SNR_a) \\ \text{Subject to}: & \quad b1) \; P_{Inter-cell}(\xi) \leq I_{Th,b}, \; b=1,...,A, \; b \neq a \\ & \quad b2) \; P_{transmit}: Tr\{\tilde{\mathbf{b}}_a\} \leq P_{Th} \\ & \quad b3) \; \tilde{\mathbf{b}}_a \geq 0, \; a=1,...,A. \end{aligned} \quad (4)$$

The Ergodic capacity as the cost function related to the $l$th channel for the $a$th cell should be maximized that can be written as (4.a) with respect to [13-eq. 13]. $A_{R_a}$ is rain attenuations in dB. (4.b.1) implies a threshold for the $a$th cell that should be defined for the worst case of the interference power over the $b$th cell caused by the $a$th cell as $I_{Th,b}$, where $b=1,...,A, b \neq a$. It should be noted that, since no knowledge of neighbor cells is available, the maximum of the mentioned power as its worst case should be explored. Defining a threshold for this maximum, the optimum solution can be robustly satisfied. (4.b.2) specifies the threshold $P_{Th}$ for the transmit-power $\|\mathbf{B}_a\|_F^2$, while the third constraint as (4.b.3) determines that $\tilde{\mathbf{b}}_a$ should be *positive semi-definite*.

The used methodology is given as follows. According to the following theorem, a closed-form solution for the considered problem can be realized. This is accomplished using the Lagrange dual-decomposition method under the assumption that the optimization problem is *convex*. The convexity is not discussed in the paper.

**Theorem 1**: *The optimum solution $\tilde{\mathbf{b}}_a^{*+}$ in the $a$th cell can be obtained by solving (5) in which $\mu_1$ and $\mu_2$ are the non-negative Lagrange multipliers associated with (4.b.1) and (4.b.2). (5) can be written as (proof: see Appendix C):*

$$\frac{\log_2 e \times \{10^{\frac{-A_{R_a}}{10}} \mathbf{D}_a^H\}}{\{1+10^{\frac{-10 A_{R_a}}{10}} Tr\{\tilde{\mathbf{b}}_a^{*+} \mathbf{D}_a\}\}} = \mu_1 \left\{ \tilde{\mathbf{H}}_{2,a}^H \tilde{\mathbf{H}}_{2,a} + \sqrt{M}\xi(\sqrt{M}\xi + 2\|\tilde{\mathbf{H}}_{2,a}\|_F)\mathbf{I}_M \right\}^H + \mu_2. \quad (5)$$

Now, for flat un-correlated Suzuki fading channels, $\mathbf{D}$ is changed as $M(\beta + \alpha)\mathbf{I}_M$. See Appendix D.

## IV. ANALYTICAL RESULTS

As a default, some of the most important considered parameters are listed as follows. BPSK modulation is used in the simulations, while neither source coding nor channel coding. $m$ related to Nakagami-$m$ fading channels is considered 0.8. The input bits are produced in a way that they are normalized, thus, the transmit-power is set to 1 Watt.

**Algorithm 1** Dogleg method
0: Given $\mu_1(0)$ and $\mu_2(0)$ as the knees of the initial doglegs $\theta_{\mu_1}(0)$ and $\theta_{\mu_2}(0)$ associated with $\mu_1^*$ and $\mu_2^*$.
1: Set q=0.
2: Repeat.
3: Calculate the optimum solution $\tilde{\mathbf{b}}_a^{*+}$ solving (6).
4: Update $\theta_{\mu_1}(t+1)$ and $\theta_{\mu_2}(t+1)$ as the new doglegs, using $\mu_1(t)$ and $\mu_2(t)$, $\theta_{\mu_1}(t)$ and $\theta_{\mu_2}(t)$. (See e.g. [14]).
5: q ← q+1
6: The dogleg method is not guaranteed: Stop processing

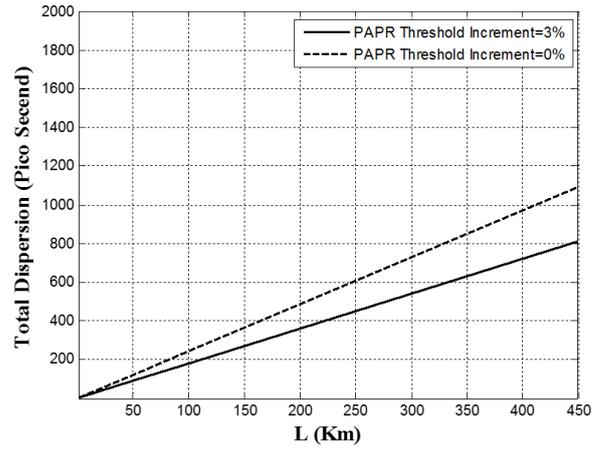

Fig.5. The total dispersion versus the cable length, while changing the threshold of PAPR

Hence, $P_{Th}$ is set to the SNR regime ([8]) in the range [0 dB, 15 dB].

This SNR is obtained after applying rain attenuations ($A_{R_a}$). $M$ is set to 2. $I_{Th,b}$ is set to 100mWatt as the 0.1 of the minimum amount of SNR (similar to [8]). Meanwhile, $a$ as the number of cells is set to 2, but this is varied in Fig. 6. Finally, the threshold $\xi$ related to the channel estimation error is set to 0, but this is varied in Fig.6. In other words, the system performance is checked out, while changing these parameters to see that how the performance is affected. The dogleg method that is used to solve the optimization problem (b), is given.

The total dispersion versus the cable length as $L$ is shown in Fig. 5, while changing $PAPR_{th}$. As this threshold is increased, the total dispersion is improved. Simultaneously, the total dispersion is linearly varied, while changing $L$. As observed, the considered trade-off between PAPR and the total dispersion for the cable length regime is acceptable. It is apparent from the figure that the increase in PAPR is 0.13 dB, whereas the total dispersion for e.g. 100

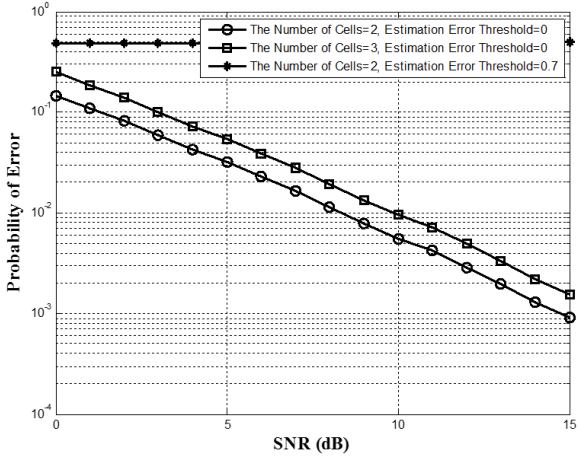

Fig.6. The probability of error versus SNR, while changing the number of cells, meanwhile changing the estimation error threshold

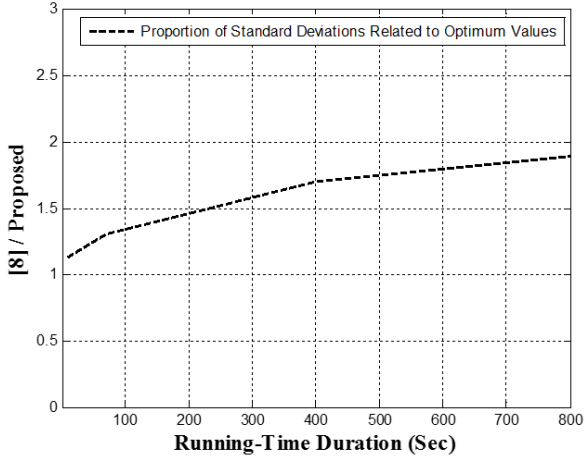

Fig.7. The standard deviation for the optimum values in this study compared to [8], during execution times

Km is approximately improved as 1.3 dB. Thus, this scheme can be logically trusted. To this end, we can justify that $PAPR_{th}$ is increased to have correspondingly a smaller *search-region* in the optimization problem (a), so as to have smaller values for the objective function $t_1$.

Let us follow up the analysis topic in down-link. The total probability of network error is illustrated against SNR regime in Fig. 6, while changing the number of cells, meanwhile while changing $\xi$. This probability can be easily proven as $1-\prod_{a=1}^{A}(1-P_a)$. in here, $P_a$ is the probability of error in the $a$th cell that can be written versus SNR such as [15]. This is increased due to network crowding caused by increasing the number of cells. Meanwhile, in practice, $a$ is practically assumed larger. On the other hand, the increase of $\xi$ makes the system performance to be destroyed. In fact, the reliability of the system imposes that the error in the estimation of the channels between cells should be logical, as much as possible.

Fig. 7 takes a comparison between [8] and this work, based on the standard deviation related to the optimum values for the optimization problem (b) and the assumed optimization problem in [8]. Using dogleg method, the mentioned standard deviation in here is smaller than [8], during execution times. In fact, the obtained closed-form solution makes the smaller probability of outgoing from *trust-region* for the proposed system. I.e., the optimum values related to the optimization problem (b) can be converged better than [8]. The larger times indicate more complexities.

## V. CONCLUSION

Three main solutions for MIMO CDMA-based optical satellite links were investigated as novelties. Two design problems were described. The first problem as the optimization problem (a) was formulated to reduce dispersions over fiber cables, as much as possible. The optimization problem (b) was adjusted, while robust solving in the multi-cell scenario over flat un-correlated Nakagami-*m*/Suzuki fading channels due to non-cooperative scheme as ICSI. System performance was investigated, considering some criteria.

## APPENDIX

### APPENDIX A. HOW TO OBTAIN SNR

$$SNR = \frac{1}{\partial^2_{N_a}}\left\{\|\mathbf{I}_M \otimes \mathbf{H}_{1,a}vec(\mathbf{B}_a)\|_F^2 = Tr\left\{\mathbf{b}_a\mathbf{b}_a^H \mathbf{H}_{1,a}^H \mathbf{H}_{1,a}\right\}\right\}, \quad (A-1)$$

Slightly similar to [8-eq. 4] on the one hand, and regarding (A-1) on the other hand, SNR can be derived, where (A-2) is satisfied, while $\tilde{\mathbf{b}}_a \triangleq$ . $\partial^2_{N_a}$ is the variance of the sky noise matrix $\mathbf{N}_a$ in the introduced satellite link for the $a$th cell. Meanwhile, it is assumed that the elements of channel matrices equally have the means *Mean* as $\sqrt{\alpha}$ and the variances *Var* as $\beta$.

$$\mathbf{D}_{M \times M} \triangleq \quad _{,a}\mathbf{H}_{1,a}^H \} = \begin{bmatrix} h_{11} & .. & h_{1M} \\ .. & .. & .. \\ h_{M1} & .. & h_{MM} \end{bmatrix}\begin{bmatrix} h_{11}^* & .. & h_{M1}^* \\ .. & .. & .. \\ h_{1M}^* & .. & h_{MM}^* \end{bmatrix}$$

$$= M\begin{bmatrix} \beta+\alpha & \alpha & \alpha \\ \alpha & .. & \alpha \\ \alpha & \alpha & \beta+\alpha \end{bmatrix}, h_{ij}h_{pq}^* = \begin{cases} \beta+\alpha, & \text{ij=pq} \\ \alpha, & \text{ij} \neq \text{pq} \end{cases},$$

(A-2)

### APPENDIX B. PROOF OF PROPOSITION 1

The maximum of the inter-cell interference as its worst case related to the *2*th channel in the *a*th cell can be stated, where (B-1), (B-2), (B-3) and (B-4) are respectively satisfied as follows:

$$\| \mathbf{P} + \mathbf{Q} \|_F \leq \| \mathbf{P} \|_F + \| \mathbf{Q} \|_F, \quad \text{(B-1)}$$

and

$$\max_{\|\Delta\|_F \leq \xi} \| \tilde{\Delta} \|_F \stackrel{[33-eq.541]}{=} \sqrt{Tr\{(\Delta \otimes \mathbf{I}_M)(\Delta^H \otimes \mathbf{I}_M^H)\}} = \{Tr\{\begin{bmatrix} [\Delta] & [0] & [0] \\ [0] & .. & [0] \\ [0] & [0] & [\Delta] \end{bmatrix}$$

$$\begin{bmatrix} [\Delta]^H & [0] & [0] \\ [0] & .. & [0] \\ [0] & [0] & [\Delta]^H \end{bmatrix}\}\}^{\frac{1}{2}} = \sqrt{MTr\{\Delta\Delta^H\}} = \sqrt{M} \| \Delta \|_F \leq \sqrt{M}\xi, \quad \text{(B-2)}$$

$$\| \mathbf{PQ} \|_F \leq \| \mathbf{P} \|_F \| \mathbf{Q} \|_F, \quad \text{(B-3)}$$

and

$$\| \mathbf{I}_M \otimes (\hat{\tilde{\mathbf{H}}}_{2,a} + \Delta) vec(\mathbf{B}_a) \|_F^2$$
$$\stackrel{(B-1)}{\leq} \{\| \tilde{\mathbf{H}}_{2,a} \mathbf{b}_a \|_F + \| \tilde{\Delta} \mathbf{b}_a \|_F\}^2$$
$$\stackrel{(B-2,3)}{\leq} \{\{\| \tilde{\mathbf{H}}_{2,a} \mathbf{b}_a \|_F + \sqrt{M}\xi \| \mathbf{b}_a \|_F\}^2 =$$
$$\| \tilde{\mathbf{H}}_{2,a} \mathbf{b}_a \|_F^2 + M\xi^2 \| \mathbf{b}_a \|_F^2 + 2\sqrt{M}\xi \| \tilde{\mathbf{H}}_{2,a} \mathbf{b}_a \|_F \times \| \mathbf{b}_a \|_F\}$$
$$\stackrel{(B-3)}{\leq} \{\| \tilde{\mathbf{H}}_{2,a} \|_F^2 + M\xi^2 + 2\sqrt{M}\xi \| \tilde{\mathbf{H}}_{2,a} \|_F\} \times \| \mathbf{b}_a \|_F^2. \quad \text{(B-4)}$$

$[0]$ in (B-2) is the all zeros matrix of size $M \times M$. (B-4) is written non-similar to [8-eqs. 6-8]. $\tilde{\Delta}$ is $\Delta \otimes \mathbf{I}_M$, while $\Delta$ is the error matrix of size $M \times M$. (B-1) and (B-3) are written according to two indispensable matrix norm features, namely the *triangle inequality* and the *Cauchy-Schwartz inequality*. Note that (4) is achieved from (B-4). Now, when this worst case is constrained, the optimum solution is completely satisfied.

It is evident that the other evaluation of the triangle inequality is examined, namely the *sub-adittivity*, whereas in [8], $\| \mathbf{P} - \mathbf{Q} \|_F \geq \| \mathbf{P} \|_F - \| \mathbf{Q} \|_F$ was used for the worst case of SNR as the cost function. Meanwhile, $\| \mathbf{b}_a \|_F^2 = Tr\{\tilde{\mathbf{b}}_a\}$.

APPENDIX C. PROOF OF THEOREM 1

Writing the *Lagrange function* as:

$$Lagrange(\tilde{\mathbf{b}}_a^{*+}, \mu_1, \mu_2) = \text{Cost-Function}(\tilde{\mathbf{b}}_a^{*+}) +$$
$$\mu_1\{P_{Inter-cell}(\xi, \tilde{\mathbf{b}}_a^{*+}) - I_{Th,b}\} + \mu_2\{P_{transmit}(\tilde{\mathbf{b}}_a^{*+}) - P_{Th}\}, \quad \text{(C-1)}$$

knowing $\frac{\delta Tr\{\mathbf{Ab}\}}{\delta \mathbf{b}} = \mathbf{A}^H$ as [16-eq.100], applying the *Karush-Kuhn-Tucker conditions* and knowing that the derivative of $log_2(1+ax)$ is $\{a/(1+ax)\}log_2 e$, (6) can be precisely achieved. As completely discussed before, this is fulfilled, assuming that the optimization problem (b) is convex. Therefore, the Lagrange dual-decomposition is exploited. However, while using this method, it is necessary to investigate the convexity of the optimization problem (b). The convexity in here can be easily studied.

APPENDIX D. SUZUKI FADING CHANNELS

A Suzuki channel can be modeled as a Rayleigh channel that is mathematically multiplied by a Log-normal channel ([17]). Therefore, **D** can be easily written for Suzuki channels, regarding (D-1) as:

$$Var[Suz] = E[Log^2]E[Ray^2] - \{E[Log]\}^2\{E[Ray]\}^2 =$$
$$(Var_{Log} + Mean_{Log}^2)(Var_{Ray} + Mean_{Ray}^2) - 0 = \beta + \alpha,$$
$$Mean[Suz] = E[Log \times Ray] = Mean_{Log} \cdot \cancel{Mean_{Ray}}^0 = 0. \quad \text{(D-1)}$$

Indeed, the shadow effects related to Log-normal channels are considered as *channel impacts* as means and variances. Now, since **D** can be represented based upon an identity matrix in here, (5) can be easily decomposed to eigen-value forms, according to the eigen-value decomposition as a helpful point to design transceivers.